# О КОРРЕЛЯЦИИ СВОЙСТВ И СТЕПЕНИ УПОРЯДОЧЕННОСТИ СТРУКТУРЫ КОМПОЗИТОВ

## А. Н. Герега

Предложена компьютеризованная методика расчета относительной степени упорядоченности структуры композиционных материалов. Показано наличие корреляции между упорядоченностью структуры и уровнем свойств материала. Обсуждается возможность уточнения терминологии, использующейся при описании структуры.

Proposed the computerized method for calculating the relative level of order composites. Correlation between a level of structure order and properties of solids is shown. Discussed the possibility of clarifying the terminology used in describing the structure.

Ключевые слова: структура, уровень свойств, упорядоченность, энтропия, функционал Ляпунова.

Памяти Ю. Л. Климонтовича

1. Введение

В современном естествознании понятие структуры относится к числу базовых. Это проявляется, в частности, в отсутствии исчерпывающего определения структуры в конечных дефинициях: структура как фундаментальное понятие определяется списком свойств. В этом перечне присутствуют и количественные величины (постоянные решетки, диаметры атомов, координационные числа, энергии связи и другие), и качественные характеристики (отмечается наличие или отсутствие дальнего порядка, определяется вид решетки, тип симметрии и другое).

Многочисленные исследования, посвященные конкретизации зависимости структуры и свойств, как правило, содержат фотографии поверхностей или срезов образцов, делающих наглядным различие структуры композитов с отличающимися свойствами. Осознавая важность зависимости свойств материалов от структуры, понимая, что, по сути, свойства и есть проявление структуры, авторы многих статей приводят вербальное описание фотографий, используя при этом термины интуитивно понятные – структурированность, неоднородность, сложность, – но не имеющие четких определений, алгоритмов расчета, количественной составляющей.

В этой связи уместно иметь характеристику, которая позволила бы количественно оценивать по фотографиям уровень упорядоченности структуры. Одной из величин, дающих такую возможность, является относительная степень упорядоченности структуры вещества, определяемая по изображениям [1], – количественная характеристика, которую рассчитывают по операбельному алгоритму, базирующемуся на представлении об энтропии информационной системы.

2. Относительная степень упорядоченности изображений

Рассмотрим вопрос о сравнении двух изображений как информационных объектов, и определим относительную степень их упорядоченности (ОСУ) [1].

Смысловое и визуальное содержание изображения существенно зависит от особенностей наблюдателя, его психологической установки, способа рассмотрения и других обстоятельств [2,3], формирование же объективных характеристик, способов их определения и измерения требует, прежде всего, формализации изображения [4-6]. Кроме этого, при определении относительной характеристики с необходимостью встает вопрос о корректности сравнения. В случае изображений адекватное сопоставление возможно по аналогии с тем, как оно вводится в теореме Гиббса об энтропии равновесного и произвольного состояний [7-9].

Энтропия Гиббса, как известно, определяется средним значением логарифма функции равновесного распределения полного набора координат и импульсов частиц системы. Сравнение значений энтропии равновесного и неравновесного состояний может производиться только при одинаковых значениях средних энергий, тем самым ограничивая набор неравновесных состояний. Условия нормировки равновесной и неравновесной функции распределения также должны быть одинаковы [7-9].

Представим компьютерное изображение как массив пикселей, выполненных в 256 оттенках серого цвета. В применении к изображениям, выполненным в оттенках, одинаковым в сравниваемых изображениях должно быть среднее количество уровня серого (УС) на пиксель. Для выполнения этого условия по данным массивов восстанавливаются функции плотности распределения значений уровня серого $f_1(i)$ и $f_2(i)$ в пикселях изображений, затем по формуле

$$\breve{f}_2(i) = f_2(i) \left[ \sum_{i=0}^{255} f_1(i) \Big/ \sum_{i=0}^{255} f_2(i) \right] \qquad (1)$$

проводится перенормировка одной из них.

В [1] показано (I-теорема), что мерой относительной степени упорядоченности двух равновеликих изображений с одинаковым значением среднего уровня серого есть функционал Ляпунова

$$S_1 - \breve{S}_2 = -\sum_{i=0}^{255}\left[f_1(i)\ln f_1(i) - \breve{f}_2(i)\ln \breve{f}_2(i)\right] = -\sum_{i=0}^{255}\left[f_1(i)\left(\ln f_1(i)/\breve{f}_2(i)\right)\right]. \tag{2}$$

Заметим, что аналогичный результат известен в теории информации (расстояние Кульбака-Лейблера – мера различия вероятностных распределений [10, 11] ), и в теории открытых систем (S-теорема Климонтовича – критерий относительной степени упорядоченности неравновесных состояний открытых систем [12, 13]).

3. Компьютерная реализация методики

Изображение, представленное в оттенках серого цвета, программно преобразуется в одномерный массив данных, который содержит численные значения уровня серого в каждом пикселе исходного изображения. При формировании массива в качестве правила обхода изображения используется бустрофедон (греч. «путь быка» – тип письма с переменным направлением строк: нечётные строки пишутся справа налево, чётные – наоборот) [14]. Это обусловлено тем, что возможные варианты оценки относительной степени упорядоченности, предусмотренные в компьютерной программе, связаны не только с анализом отдельных пикселей, но и со сравнением УС соседних пикселей: по разности или отношению, по абсолютной величине этой разности и другим.

Алгоритм выравнивания интегрального УС базируется, как отмечалось, на перенормировке функции плотности распределения серого в пикселях. Одно из сравниваемых изображений выбирается в качестве эталонного, и для достижения равного УС либо проводят растяжения-сжатия функции распределения второго изображения по формуле (1), либо ее сдвиг по оси, на которой откладывается значение уровня серого. Если существуют требования, минимизирующие возможности модификации изображения, то для достижения равного среднего УС функции распределения обоих изображений смещаются в противоположных направлениях [1].

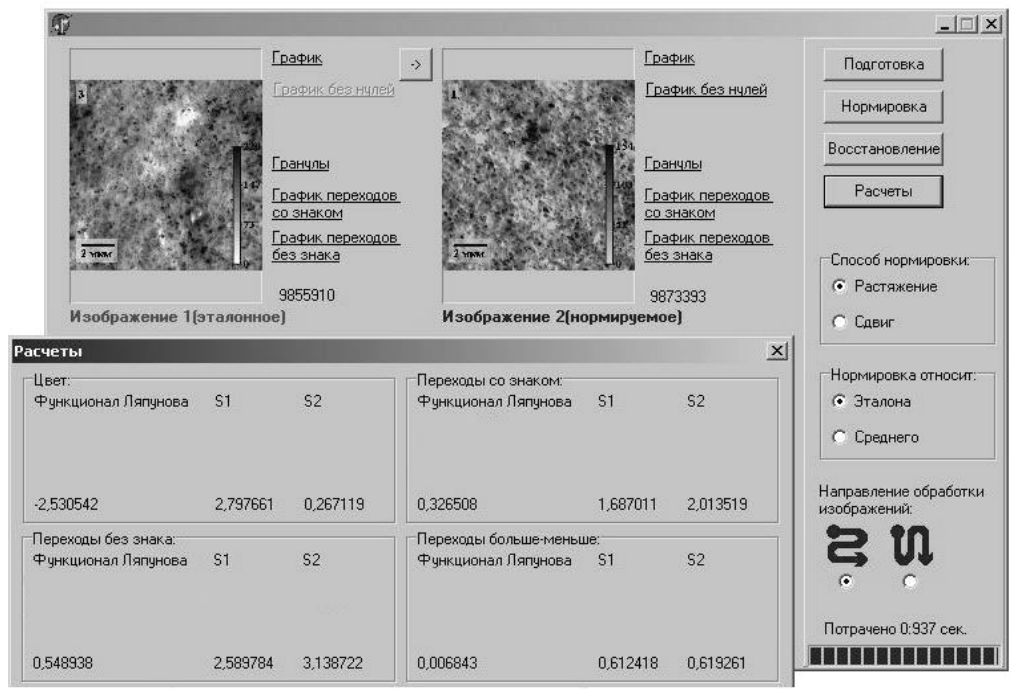

Рис.1. Главный интерфейс программы.

С учетом возможностей программы по варьированию режимов обработки массивов (см. рис. 1), для каждой пары сравниваемых изображений можно получить по 32 результата вычисления функционала Ляпунова. По нашим данным наиболее чувствительными, как правило, являются оценки, сделанные по величине уровня серого в пикселях.

4. Два примера

В [15] методами интерференционной микроскопии проведено исследование структуры и свойств оксидных покрытий, нанесенных на поверхность деформируемого алюминиевого сплава В95Т1, обычно используемого для изготовления работающих на сжатие конструкций. В работе экспериментально получены значения некоторых параметров образцов. Установлено, в частности, что полученные по различным техно-логиям покрытия, имеют микротвердость и модуль упругости, отличающиеся соответственно на 15 и 8 % (табл. 1).

По нашим данным относительная степень упорядоченности, рассчитанная по картам интенсивности поверхности образцов (рис. 2), показывает, что образец А более упорядочен в описанном выше смысле: ОСУ равна 0,26. Это значит, что с ростом упорядоченности структуры покрытия его микротвердость возрастает, а модуль упругости уменьшается (см. табл. 1).

Табл. 1. Механические характеристики покрытий (фрагмент таблицы из работы [15]).

| Параметр | Образец А | | Образец Б | |
|---|---|---|---|---|
| | Значение | ± | Значение | ± |
| Микротвердость, ГПа | 10,65 | 1,6 | 9,26 | 2,12 |
| Модуль упругости, ГПа | 157,2 | 12,16 | 170,1 | 12,4 |

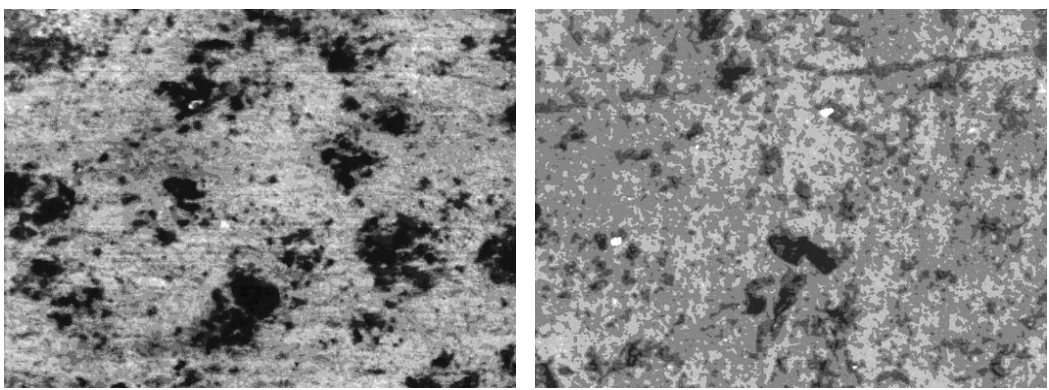

Рис. 2. Карты интенсивности поверхности образцов А и Б [15].

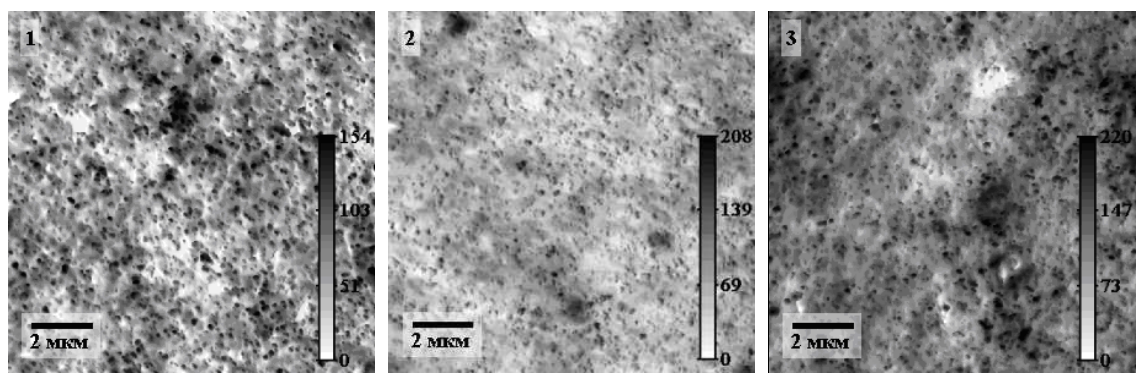

Рис. 3. Фотографии поверхности наполненной резины, полученные с помощью атомно-силового микроскопа [16].

В [16] исследована зависимость механических свойств резин, содержащих технический углерод, от «однородности распределения наполнителя в материале». Автор [16] считает, что «в образцах 1 и 3 распределение можно считать равномерным, а во втором наблюдаются скопления кластеров в отдельные группы (см. рис. 3)». Он также полагает, что «чем неоднороднее распределение и больше размеры кластеров, тем сильнее взаимодействие наполнитель-наполнитель, неоднороднее поле напряжений и больше гистерезисных потерь при циклической нагрузке. С этой точки зрения, наихудшими механическими характеристиками обладает второй образец» [16].

Использование представления об ОСУ изображений при распределении уровня серого показал, что второй образец – более упорядоченный, чем первый ($\Delta S = 4,683$) и третий ($\Delta S = 4,752$). С учетом результатов [16] видно, что с ростом упорядоченности структуры углеродных кластеров в резине ее механические свойства ухудшаются.

5. Заключение

Относительную степень упорядоченности можно рассматривать как интегральную количественную характеристику структуры физических тел, которая является альтернативой описательным понятиям типа «структурированность», «неоднородность», «сложность».

Очевидна возможность ввода в рассмотрение «абсолютной» упорядоченности. Для ее определения нужно выбрать нуль шкалы: это может быть упорядоченность изображения, созданного генератором случайных чисел с равномерным распределением, или, напротив, – полная упорядоченность «черного квадрата».

Литература